\documentclass[conference]{IEEEtran}
\usepackage{multirow}
\usepackage[ruled,vlined]{algorithm2e}
\usepackage{amssymb}
\usepackage{mathtools} 
\usepackage{url}
\usepackage{blindtext, graphicx}
\usepackage{graphicx}
\usepackage{caption}
\usepackage{tabularx}
\usepackage{subfigure}
\usepackage{multirow}
\usepackage{amsmath}
\usepackage{indentfirst}
\usepackage{parskip}
\usepackage[ruled,vlined]{algorithm2e}
\ifCLASSINFOpdf
\else
\fi 
\begin{document}
\title{Investigating the Application of Common-Sense Knowledge-Base for Identifying Term Obfuscation in Adversarial Communication}
\author{\IEEEauthorblockN{Swati Agarwal}
\IEEEauthorblockA{Indraprastha Institute of Information Technology\\
New Delhi, India\\
Email: swatia@iiitd.ac.in}
\and
\IEEEauthorblockN{Ashish Sureka}
\IEEEauthorblockA{ABB Corporate Research\\
Bangalore, India\\
Email: ashish.sureka@in.abb.com}}
\maketitle
\begin{abstract}
Word obfuscation or substitution means replacing one word with another word in a sentence to conceal the textual content or communication. Word obfuscation is used in adversarial communication by terrorist or criminals for conveying their messages without getting red-flagged by security and intelligence agencies intercepting or scanning messages (such as emails and telephone conversations). ConceptNet is a freely available semantic network represented as a directed graph consisting of nodes as concepts and edges as assertions of common sense about these concepts. We present a solution approach exploiting vast amount of semantic knowledge in ConceptNet for addressing the technically challenging problem of word substitution in adversarial communication. We frame the given problem as a textual reasoning and context inference task and utilize ConceptNet's natural-language-processing tool-kit for determining word substitution. We use ConceptNet to compute the conceptual similarity between any two given terms and define a Mean Average Conceptual Similarity (MACS) metric to identify out-of-context terms. The test-bed to evaluate our proposed approach consists of Enron email dataset (having over $600000$ emails generated by $158$ employees of Enron Corporation) and Brown corpus (totaling about a million words drawn from a wide variety of sources). We implement word substitution techniques used by previous researches to generate a test dataset. We conduct a series of experiments consisting of word substitution methods used in the past to evaluate our approach. Experimental results reveal that the proposed approach is effective.
\end{abstract}
\IEEEpeerreviewmaketitle
\section{Research Motivation and Aim}
Intelligence and security agencies intercepts and scans billions of messages and communications every day to identify dangerous communications between terrorists and criminals. Surveillance by Intelligence agencies consists of intercepting mail, mobile phone and satellite communications. Message interception to detect harmful communication is not only done by Intelligence agencies to counter terrorism but also by law enforcement agencies to combat criminal and illicit acts for example by drug cartels or by organizations to counter employee collusion and plot against the company. Law enforcement and Intelligence agencies have a watch-list or lexicon of red-flagged terms such as attack, bomb and heroin. The watch-list of suspicious terms are used for keyword-spotting in intercepted messages which are filtered for further analysis \cite{Deshmukh14}\cite{Fong08}\cite{Fong06}\cite{Roussinov07}\cite{Jabbari08}. \\
\indent Terrorist and criminals use textual or word obfuscation to prevent their messages from getting intercepted by the law enforcement agencies. Textual or word substitution consists of replacing a red-flagged term (which is likely to be present in the watch-list) with an "ordinary" or an "innocuous" term. Innocuous terms are those terms which are less likely to attract attention of security agencies. For example, the word \textsf{attack} being replaced by the phrase \textsf{birthday function} and \textsf{bomb} being replaced by the term \textsf{milk}. Research shows that terrorist use low-tech word substitution than encryption as encrypting messages itself attracts attention. Al-Qaeda used the term \textsf{wedding} for \textsf{attack} and \textsf{architecture} for World Trade Center in their email communication. Automatic word obfuscation detection is natural language processing problem that has attracted several researcher's attention. The task consists of detecting if a given sentence has been obfuscated and which term(s) in the sentence has been substituted. The research problem is intellectually challenging and non-trivial as natural language can be vast and ambiguous (due to polysemy and synonymy) \cite{Deshmukh14}\cite{Fong08}\cite{Fong06}\cite{Roussinov07}\cite{Jabbari08}. \\
\indent ConceptNet\footnote{\url{http://conceptnet5.media.mit.edu/}} is a semantic network consisting of nodes representing concepts and edges representing relations between the concepts. ConceptNet is a freely available commonsense knowledgebase ehich contains everyday basic knowledge \cite{havasi2007conceptnet}\cite{liu2004conceptnet}\cite{speer2013conceptnet}. It has been used as a lexical resource and natural language processing toolkit for solving many natural language processing and textual reasoning tasks \cite{havasi2007conceptnet}\cite{liu2004conceptnet}\cite{speer2013conceptnet}. We hypothesize that ConceptNet can be used as a semantic knowledge-base to solve the problem of textual or word obfuscation. We believe that the relations between concepts in ConceptNet can be exploited to find conceptual similarity between given concepts and use to detect out-of-context terms or terms which typically do not co-occur together in everyday communication. The research aim of the study presented in the following:
\begin{enumerate}
\item To investigate the application of a commonsense knowledge-base such as ConceptNet for solving the problem of word or textual obfuscation. 
\item To conduct an empirical analysis on large and real-word datasets for the purpose of evaluating the effectiveness of the application of ConceptNet (a lexical resource to compute conceptual or semantic similarity between two given terms) for the task of word obfuscation detection. 
\end{enumerate}
\begin {table}[t]
\sffamily\normalsize
\caption {\textbf{List of Previous Work (Sorted in Reverse Chronological Order) in the Area of Detecting Word Obfuscation in Adversarial Communication. ED: Evaluation Dataset, RS: Resources Used in Solution Approach, SA: Solution Approach}} \label{tab:relwork} 
\begin{center}
\begin{tabular}{cp{7.5cm}}
\multicolumn{2}{l}{Deshmukh et al. 2008 \cite{Deshmukh14}} \\ \hline
ED & Google News \\
RS& Google search engine \\
SA & Measuring sentence oddity, enhance sentence oddity and k-grams frequencies \\ 
&\\
\multicolumn{2}{l}{Jabbari et al. 2008 \cite{Jabbari08}} \\ \hline
ED& British National Corpus (BNC) \\ 
RS& 1.4 billion words of English Gigaword v.1 (newswire corpus) \\
SA& Probabilistic or distributional model of context \\
&\\
\multicolumn{2}{l}{Fong et al. 2008 \cite{Fong08}} \\ \hline
ED & Enron e-mail dataset, Brown corpus \\ 
RS& British National Corpus (BNC), WordNet, Yahoo, Google and MSN search engine \\
SA & Sentence oddity, K-gram frequencies, Hypernym Oddity (HO) and Pointwise Mutual Information (PMI) \\ 
&\\
\multicolumn{2}{l}{Fong et al. 2006 \cite{Fong06}} \\ \hline
ED & Enron e-mail dataset \\
RS & British National Corpus (BNC), WordNet, Google search engine \\ 
SA & Sentence oddity measures, semantic measure using WordNet, and frequency count of the bigrams around the target word \\ 
\end{tabular}
\end{center}
\end{table}
\section{Background}
In this Section, we discuss closely related work to the study presented in this paper and explicitly state the novel contributions of our work in context to previous researches. Term obfuscation in adversarial communication is an area that has attracted several researcher's attention. Table \ref{tab:relwork} displays list of traditional techniques sorted in reverse chronological order of their publication. Table \ref{tab:relwork} shows the evaluation dataset, lexical resource and the high level solution approach applied in each of the four techniques. The solution approaches consists of measuring sentence oddity using results from Google search engine, using probabilistic or distributional model of context and using WordNet and Pointwise Mutual Information (PMI) to compute out-of-context terms in a given sentence. \\
\indent ConceptNet has been used by several researchers for solving a variety of natural language processing problems. We briefly discuss some of the recent and related work. Wu et al. use relation selection to improve value propagation in a ConceptNet-based sentiment dictionary (sentiment polarity classification task) \cite{wu2014}. Bouchoucha et al. use ConceptNet as an external resource for query expansion \cite{bouchoucha2014}. Revathi et al. present an approach for similarity based video annotation utilizing commonsense knowledge-base. They apply Local binary pattern (LBP) and commonsense knowledgebase to reduce the semantic gap for non-domain specific videos automatically \cite{revathi}. Poria et al. propose a ConceptNet-based semantic parser that deconstructs natural language text into concepts based on the dependency relation between clauses. Their approach is domain-independent and is able to extract concepts from heterogeneous text \cite{poria2014}.\\
\section{Research Contributions}
In context to existing work, the study presented in this paper makes the several unique and novel research contributions:
\begin{enumerate}
\item The study presented in this paper is the first focused research investigation on the application of ConceptNet common sense knowledge-base for solving the problem of textual or term obfuscation. While there has been work done in the area of using a corpus as a lexical resource for the task of term obfuscation detection, the application of an ontology like ConceptNet for determining conceptual similarity between given terms and identifying out-of-context or odd terms in a given sentence is novel in context to previous work.
\item We conduct an in-depth empirical analysis to examine the effectiveness of the proposed approach. The test dataset consists of examples extracted from research papers on term obfuscation, Enron email dataset (having over $600000$ emails generated by $158$ employees of Enron Corporation) and Brown corpus (totaling about a million words drawn from a wide variety of sources).
\item The study presented in this paper is an extended version of our work Agarwal et al. accepted in Future Information Security Workshop Co-located with COMSNETS conference \cite{agarwal2015using}. Due to the small page limit for regular/full papers (at most six pages) in FIS, COMSNETS $2015$\footnote{\url{http://www.comsnets.org/archive/2015/fis_workshop.html}}, several aspects including results and details of proposed approach are not covered. This paper presents the complete and detailed description of our work on term obfuscation detection in adversarial communication.
\end{enumerate} 
\begin{figure}[t]
\centerline{
\includegraphics[width=0.5\textwidth]{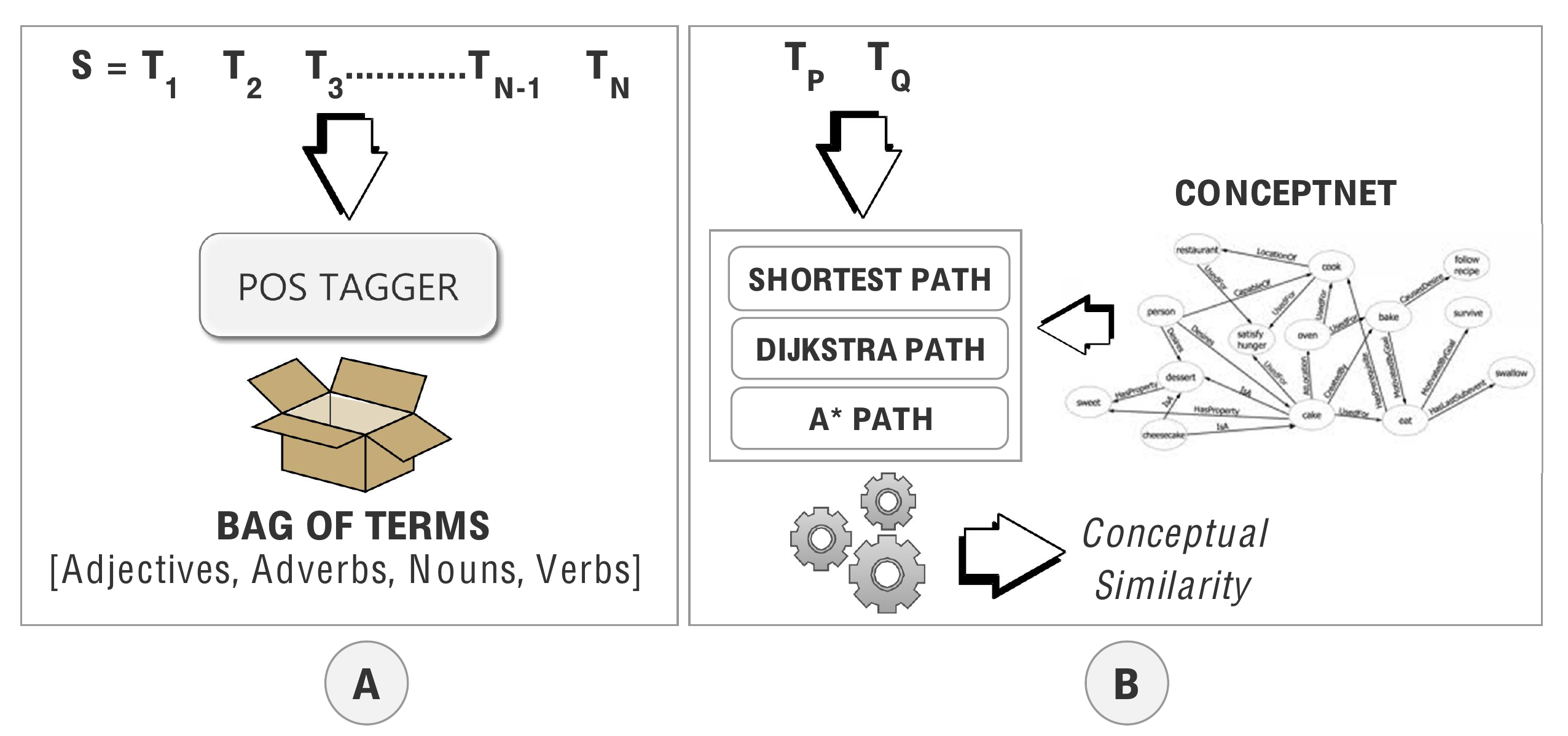}}
\caption{\textbf{Solution framework demonstrating two phases in the processing pipeline. Phase \textit{A} shows tokenizing given sentence and applying the part-of-speech-tagger. Phase \textit{B} shows computing conceptual similarity between any two given term using ConceptNet as a lexical resource and applying graph distance measures.} Source: Agarwal et al. \cite{agarwal2015using}}
\label{fig:Step-A-B}
\end{figure}
\section{Solution Approach}
\begin{figure}[t]
\centerline{
\includegraphics[width=0.5\textwidth]{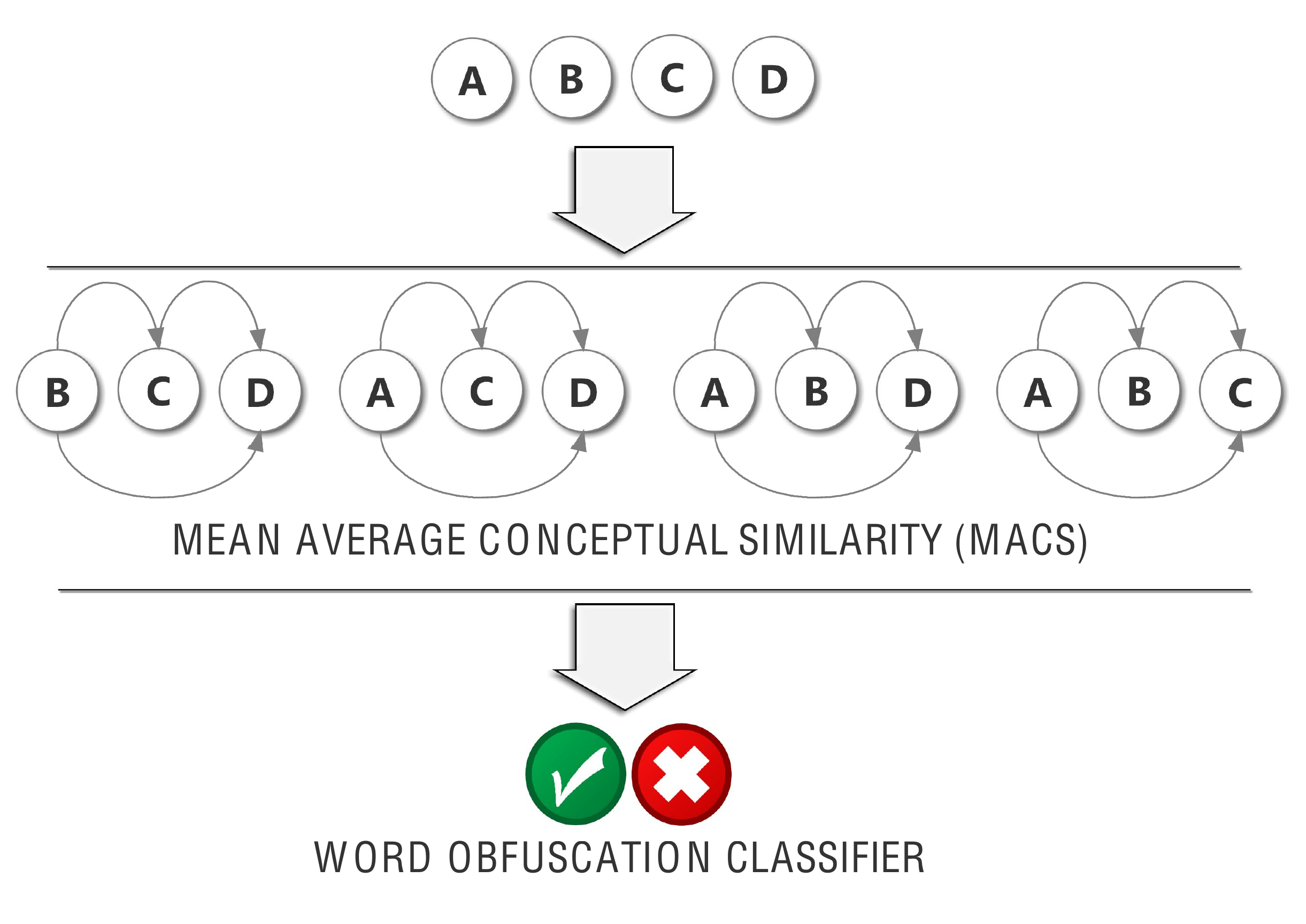}}
\caption{\textbf{Solution framework demonstrating the procedure of computing Mean Average Conceptual Similarity (MACS) score for a bag-of-terms and for determining the term which is out-of-context. The given example consisting of four terms \textit{A}, \textit{B}, \textit{C} and \textit{D} requires computing conceptual similarity between two terms $12$ times.} Source: Agarwal et al. \cite{agarwal2015using}}
\label{fig:COMPARISON}
\end{figure}

Figures \ref{fig:Step-A-B} and \ref{fig:COMPARISON} illustrates the general research framework for the proposed solution approach. The proposed solution approach primarily consists of two phases labeled as \textit{A} and \textit{B} (refer to Figure \ref{fig:Step-A-B}). In Phase \textit{A}, we tokenize a given sentence $S$ into a sequence of terms and tag each term with their part-of-speech. We use Natural Language Toolkit\footnote{\url{www.nltk.org}} (NLTK) part-of-speech tagger for tagging each term. We exclude non-content bearing terms using an exclusion list. For example, we exclude conjunctions (and, but, because), determiners (the, an, a), prepositions (on, in, at), modals (may, could, should), particles (along, away, up) and base form of verbs. We create a bag-of-terms (a set) with the remaining terms in the given sentence. As shows in Figures \ref{fig:Step-A-B} and \ref{fig:COMPARISON}, Phase \textit{B} consists of computing the Mean Average Conceptual Similarity (MACS) score for a bag-of-terms and identify obfuscated term in a sentence using the MACS score. The conceptual similarity between any two given terms $T_{p}$ and $T_{q}$ is computed by taking the average of number of edges in the shortest-path between $T_{p}$ and $T_{q}$ and the number of edges in the shortest-path between $T_{q}$ and $T_{p}$ (and hence the term \textit{average} in MACS). We use three different algorithms (Dijikstra's, A* and Shortest path) to compute the number of edges between any two given terms. The different algorithms are experimental parameters and we experiment with three different algorithms to identify the most effective algorithm for the given task. \\
Let us say that the size of the bag-of-terms after Phase \textit{A} is $N$. As shown in Figure \ref{fig:COMPARISON}, we compute the MACS score $N$ times. The number of comparisons (computing the number of edges in the shortest path) required for computing a single MACS score is twice of $^{(N-1)}P_{2}$ times. Consider the scenario in Figure \ref{fig:COMPARISON}, the MACS score is computed $4$ times for the four terms: \textit{A}, \textit{B}, \textit{C} and \textit{D}. The comparison required for computing the MACS score for \textit{A} are: $B-C$, $C-B$, $B-D$, $D-B$, $C-D$ and $D-C$. Similarly, the comparisons required for computing the MACS score for \textit{B} are: $A-C$, $C-A$, $A-D$, $D-A$, $C-D$ and $D-C$. The obfuscated term is the term for which the MACS score is the lowest. Lower number of edges between two terms indicate higher conceptual similarity. The intuition behind the proposed approach is that a term will be out of-context in a given bag-of-terms if the MACS score of terms minus the given term is low. The out-of-context term will increase the average conceptual similarity and hence the MACS score. 
\subsection{Worked-Out Example}\label{sec:woe}
We take two concrete worked-out examples in-order to explain our approach. Consider a case in which the original sentence is: "We will attack the airport with bomb". The red-flagged term in the given sentence is \textsf{bomb}. Let us say that the term \textsf{bomb} is replaced with an innocuous term \textsf{flower} and hence the obfuscated textual content is: "We will attack the airport with flower". The bag-of-terms (nouns, adjectives, adverbs and verbs and not including terms in an exclusion list) in the substituted text is [\textsf{attack}, \textsf{airport}, \textsf{flower}]. The conceptual similarity between \textsf{airport} and \textsf{flower} is $3$ as the number of edges between \textsf{airport} and \textsf{flower} is $3$ (\textsf{airport}, \textsf{city}, \textsf{person}, \textsf{flower}) and similarly, the number of edges between \textsf{flower} and \textsf{airport} is $3$ (\textsf{flower}, \textsf{be}, \textsf{time}, \textsf{airport}). The conceptual similarity between \textsf{attack} and \textsf{flower} is also $3$. The number of edges between \textsf{attack} and \textsf{flower} is $3$ (\textsf{attack}, \textsf{punch}, \textsf{hand}, \textsf{flower}) and the number of edges between \textsf{flower} and \textsf{attack} is $3$ (\textsf{flower}, \textsf{be}, \textsf{human}, \textsf{attack}). The conceptual similarity between \textsf{attack} and \textsf{airport} is $2.5$. The number of edges between \textsf{attack} and \textsf{airport} is $2$ (\textsf{attack}, \textsf{terrorist}, \textsf{airport}) and the number of edges between \textsf{airport} and \textsf{attack} is $3$ (\textsf{airport}, \textsf{airplane}, \textsf{human}, \textsf{attack}). The Mean Average Conceptual Similarity (MACS) score is $(3+3+2.5)/3$ = $2.83$. In the given example consisting of $3$ terms in the bag-of-terms, we computed the conceptual similarity between two terms six times. \\
\indent Consider another example in which the original sentence is: "Pistol will be delivered to you to shoot the president". \textsf{Pistol} is clearly the red-flagged term in the given sentence. Let us say that the term \textsf{Pistol} is replaced with an ordinary term \textsf{Pen} as a result of which the substituted sentence becomes: "Pen will be delivered to you to shoot the president". After applying part-of-speech tagging, we tag \textsf{pen} and \textsf{president} as noun and \textsf{shoot} as a verb. The bag-of-terms for the obfuscated sentence is: [\textsf{pen}, \textsf{shoot}, \textsf{president}]. The conceptual similarity between \textsf{shoot} and \textsf{president} is $2.5$ as the number of edges between \textsf{president} and \textsf{shoot} is $2$ (\textsf{president}, \textsf{person}, \textsf{shoot}) and similarly, the number of edges between \textsf{shoot} and \textsf{president} is $3$ (\textsf{shoot}, \textsf{fire}, \textsf{orange}, \textsf{president}). The conceptual similarity between \textsf{pen} and \textsf{president} is $3$. The number of edges between \textsf{president} and \textsf{pen} is $3$ (\textsf{president}, \textsf{ruler}, \textsf{line}, \textsf{pen}) and the number of edges between \textsf{pen} and \textsf{president} is $3$ (\textsf{pen}, \textsf{dog}, \textsf{person}, \textsf{president}). The conceptual similarity between \textsf{pen} and \textsf{shoot} is $3.0$. The number of edges between \textsf{shoot} and \textsf{pen} is $3$ (\textsf{shoot}, \textsf{bar}, \textsf{chair}, \textsf{pen}) and the number of edges between \textsf{pen} and \textsf{shoot} is $3$ (\textsf{pen}, \textsf{dog}, \textsf{person}, \textsf{shoot}). The Mean Average Conceptual Similarity (MACS) score is $(2.5+3+3)/3$ = $2.83$.
\begin{algorithm}
\normalsize
\SetAlgoLined\DontPrintSemicolon
\KwData{Substituted Sentence $S'$, Conceptnet Corpus $C$}
\KwResult{Obfuscated Term $O_{T}$}
\nl \For {$\textbf{all}$ record $r \in C$}{
\nl Edge $E.add (r.node_{1}, r.node_{2}, r.relation$)\;
\nl Graph $G.add ($E)\;
}
\nl $tokens=S'.tokenize()$\;
\nl $pos.add(pos\_tag(tokens)$\;
\nl \For {$\textbf{all}$ $tag \in pos$ and $token \in tokens$}{
\nl \If {tag is in (verb, noun, adjective, adverb)}{
\nl $BoW.add(token.lemma)$\;}
}
\nl \For {$iter= 0$ to BoW.length}{
\nl $concepts=BoW.pop(iter)$\;
\nl \For {$i= 0$ to concepts.length-$1$}{
\nl \For{$j=i$ to concepts.length}{
\nl \If{($i!=$j)}{
\nl path $c_{i,j}=Dijikstra_path_len(G, i, j)$\;
\nl path $c_{j,i}=Dijikstra_path_len(G, j, i)$\;
\nl avg.add(Average($c_{i,j}, c_{j,i}$)\;
}}
}
\nl mean.add(Mean(avg))\;
}
\nl $O_{T}=BoW$.valueAt(min(mean))\;
\captionsetup{justification=centering}
\caption{Obfuscated Term}\label{algo:OT}
\end{algorithm}
\begin{figure}[t]
\centerline{
\includegraphics[width=0.5\textwidth]{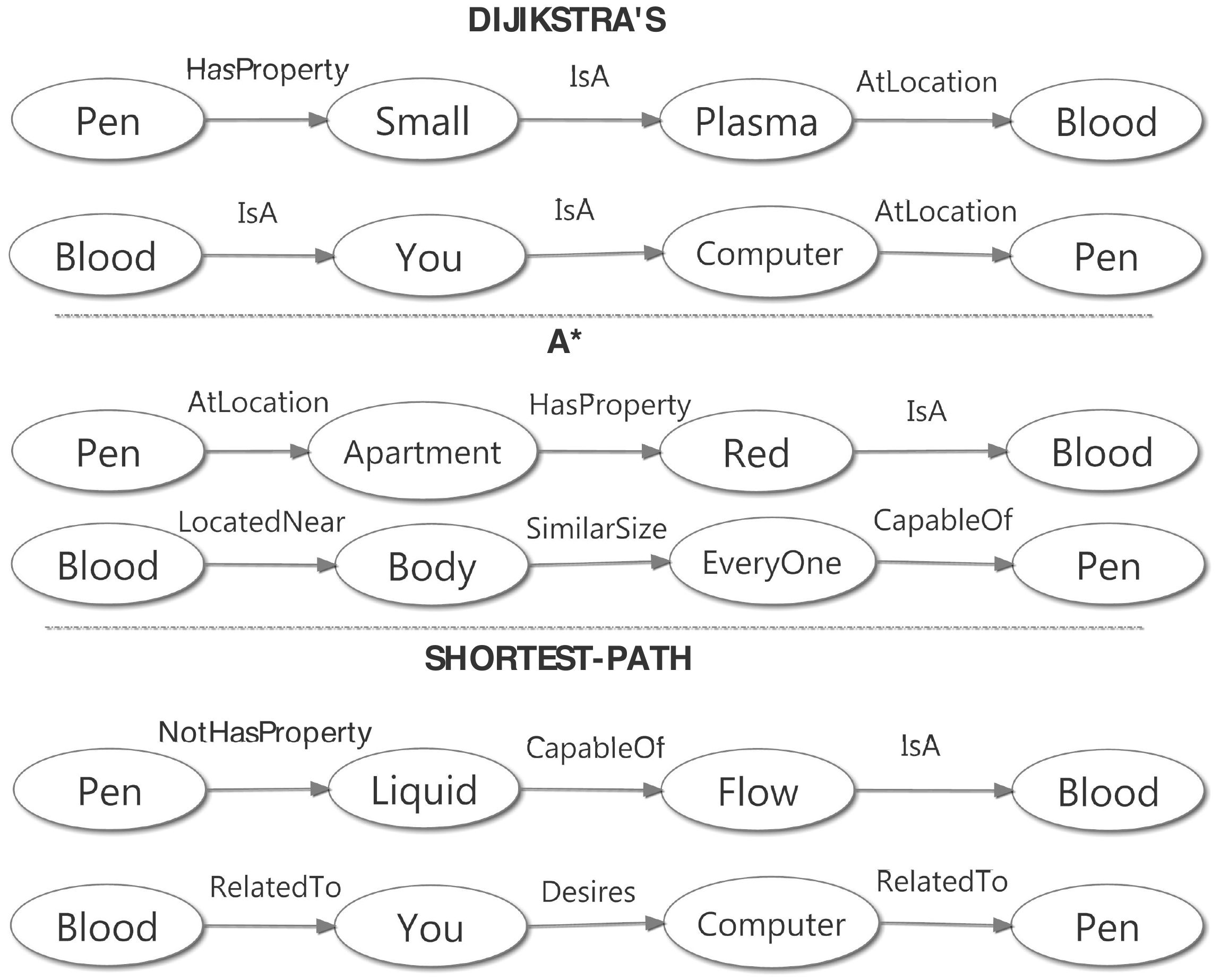}}
\caption{\textbf{ConceptNet paths (nodes and edges) between two concepts \textsf{Pen} and \textsf{Blood} using three different distance metrics}}
\label{fig:PATHS}
\end{figure}
\subsection{Solution Pseudo-code and Algorithm}
Algorithm \ref{algo:OT} describes the proposed method to identify an obfuscated term in a given sentence. Inputs to our algorithm is a substituted sentence $S'$ and the ConceptNet $4.0$ corpus $C$ (a common-sense knowledge-base). In Steps $1$ to $3$, we create a directed network graph from ConceptNet corpus where nodes represent concepts and edge represents a relation between two concepts (for example, HasA, IsA, UsedFor). As described in the research framework (refer to Figures \ref{fig:Step-A-B} and \ref{fig:COMPARISON}), in Steps $4$ to $5$, we tokenize $S'$ and apply part-of-speech tagger to classify terms according to their lexical categories (such as noun, verbs, adjectives and adverbs). In Steps $6$ to $8$, we create a bag-of-terms of the lemma of verbs, nouns, adjectives and adverbs that are present in $S'$. In Steps $9$ to $17$, we compute the mean average conceptual similarity (MACS) score for bag-of-terms. In Step $18$, we compute the minimum of all MACS scores to identify the obfuscated term. In proposed method, we use three different algorithms to compute the shortest path length between the concepts. \\
\begin{table}[t]
\centering
\sffamily
\normalsize
\caption{\textbf{Concrete Examples of Computing Conceptual Similarity between Two Given Terms Using Three Different Distance Metrics or Algorithms (NP: Denotes No-Path between the Two Terms and is Given a Default Value of $4$).} Source: Agarwal et al. \cite{agarwal2015using}} \label{tab:T1T2} 
\begin{tabular}{lllll}
\multicolumn{2}{c}{}&\multicolumn{3}{c}{\textbf{Dijikstra's Algo}}\\
Term 1 & Term 2 & T1-T2 & T2-T1 & Mean\\
\hline
Tree & Branch & 1 & 1 & 1\\
Pen & Blood & 3 & 3 & 3\\
Paper & Tree & 1 & 1 & 1\\
Airline & Pen & 4(NP) & 4 & 4\\
Bomb & Blast & 2 & 4(NP) & 3\\
\multicolumn{2}{c}{}&\multicolumn{3}{c}{\textbf{A* Algo}}\\
Tree & Branch & 1 & 1 & 1\\
Pen & Blood & 3 & 3 & 3\\
Paper & Tree & 1 & 1 & 1\\
Airline & Pen & 4(NP) & 4 & 4\\
Bomb & Blast & 2 & 4(NP) & 3\\
\multicolumn{2}{c}{}&\multicolumn{3}{c}{\textbf{BFS Algo}}\\
Tree & Branch & 1 & 1 & 1\\
Pen & Blood & 3 & 3 & 3\\
Paper & Tree & 1 & 1 & 1\\
Airline & Pen & 4(NP) & 4 & 4\\
Bomb & Blast & 2 & 4(NP) & 3\\
\end{tabular}
\end{table}
\begin{table}[t]
\centering
\sffamily
\normalsize
\caption{\textbf{Concrete Examples of Conceptually and Semantically Un-related Terms and their Path Length (PL) to Compute the Default Value for No-Path}} \label{tab:why4} 
\begin{tabular}{lll|lll}
T1 & T2 & PL & T1 & T2 & PL\\
\hline
Bowl & Mobile & 3 & Office & Festival & 3\\
Wire & Dress & 3 & Feather & Study & 3\\
Coffee & Research & 3 & Driver & Sun & 3\\
\end{tabular}
\end{table}
\indent Figure \ref{fig:PATHS} shows an example of shortest path between two terms \textsf{Pen} and \textsf{Blood} using Dijikstra's, A* and Shortest path algorithms. As shown in Figure \ref{fig:PATHS}, the path between the two terms \textsf{Pen} and \textsf{Blood} can be different than the path between the terms \textsf{Blood} and \textsf{Pen} (terms are same but the order is different). For example, the path between \textsf{Pen} and \textsf{Blood} using A* consists of \textsf{Apartment} and \textsf{Red} as intermediate nodes whereas the path between \textsf{Blood} and \textsf{Pen} consists of nodes \textsf{Body} and \textsf{EveryOne} using the A* algorithm. Also, the Figure \ref{fig:PATHS} demonstrates that the path between the same two terms is different for different algorithms.\\
Two terms are related to each other in various contexts. In ConceptNet, the path length describes the extent of semantic similarity between concepts. If two terms are conceptually similar then the path length will be smaller in comparison to the terms that are highly dissimilar. Therefore if we remove an obfuscated term from the bag-of-terms the MAC score of remaining terms will be minimum. Table \ref{tab:T1T2} shows some concrete examples of semantic similarity between two concepts. Table \ref{tab:T1T2} illustrates that the terms \textsf{Tree} \& \textsf{Branch} and \textsf{Paper} \& \textsf{Tree} are conceptually similar and has a path length of 1 which means that both the concepts are directly connected in the ConceptNet knowledge-base. $NP$ denotes no-path between the two concepts. For example, in Table \ref{tab:T1T2} we have a path length of 2 from source node \textsf{Bomb} to target node \textsf{Blast} while there is no path from \textsf{Blast} to \textsf{Bomb}. We use a default value of 4 in case of no-path between two concepts. We conduct an experiment on ConceptNet 4.0 and compute the distance between highly dissimilar terms. Table \ref{tab:why4} shows that in majority of cases the path length between semantically un-related terms is 3. Therefore we use $4$ (distance between un-related terms + $1$ for upper bound) as a default value for no-path between two concepts.
\section{Experimental Evaluation and Validation}
\begin{figure*}[t]
\centering
\begin{minipage}{.50\textwidth}
\subfigure[\label{stats_brown}Brown news corpus]{\includegraphics[width=0.49\textwidth]{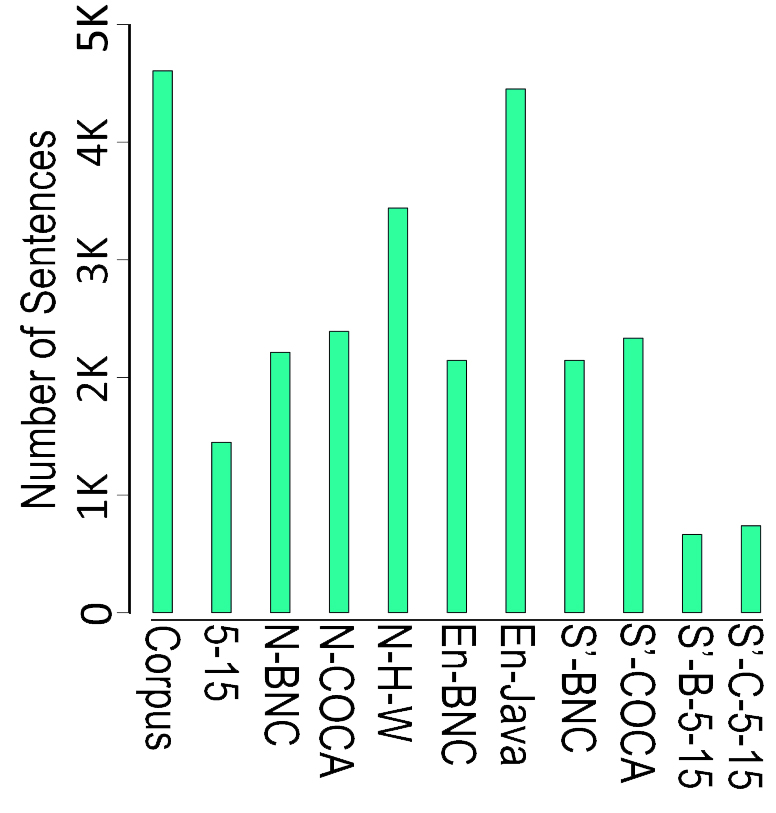}}
\subfigure[\label{stats_enron}Enron mail corpus]{\includegraphics[width=0.49\textwidth]{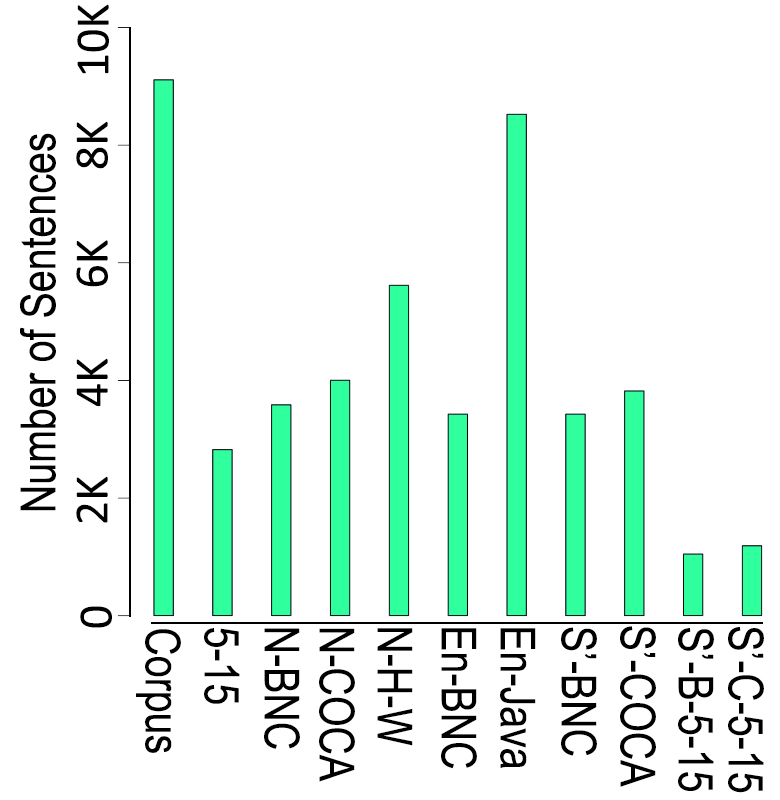}}
\caption{\textbf{Bar chart for the experimental dataset statistics (refer to Table \ref{tab:stats_brown_enron} for exact values).} Source: Agarwal et al. \cite{agarwal2015using}}
\label{dataset_stats}
\end{minipage}%
\hfill
\begin{minipage}{.24\textwidth}
 \centering
 \includegraphics[width=0.98\textwidth]{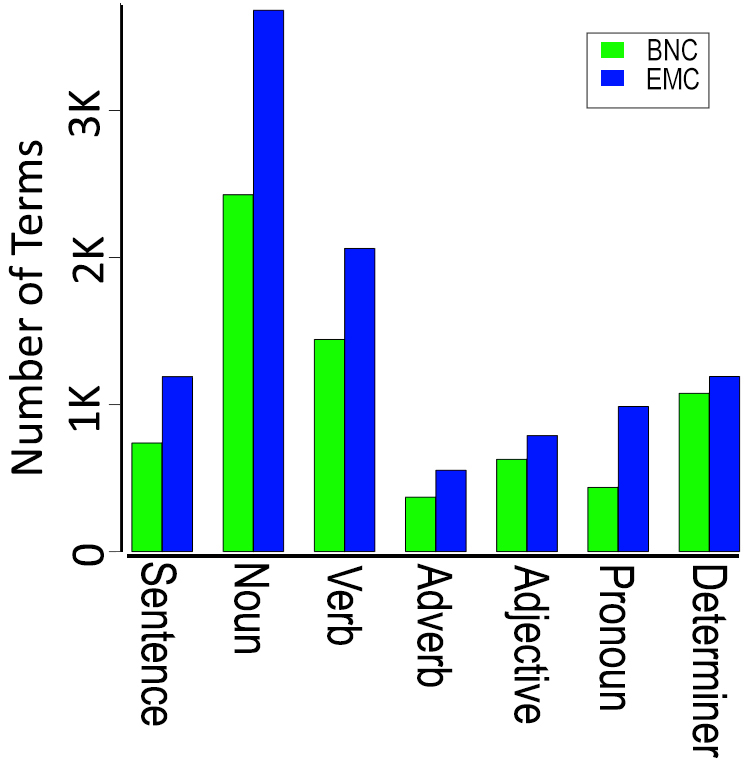}
 \caption{\textbf{Bar chart for the number of part-of-speech tags in experimental dataset.} Source: Agarwal et al. \cite{agarwal2015using}}
 \label{pos_tags}
\end{minipage}%
%
\hfill
\begin{minipage}{.24\textwidth}
 \centering
 \includegraphics[width=0.98\textwidth]{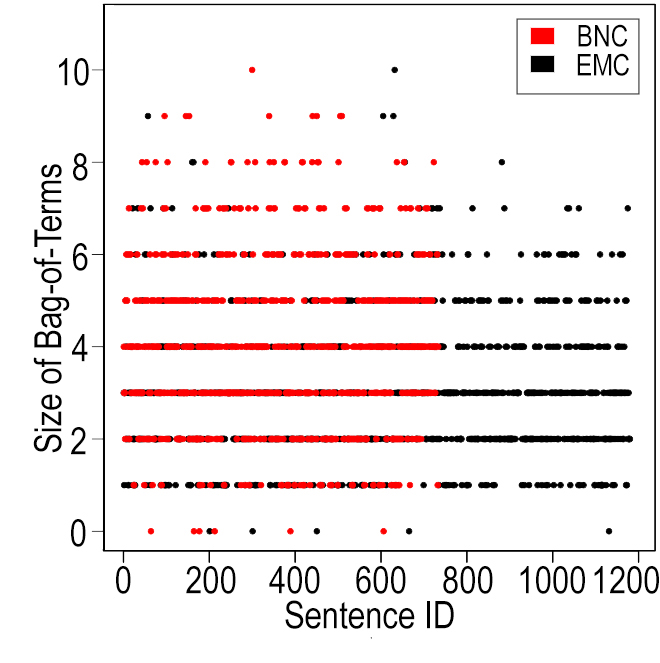}
 \caption{\textbf{Scatter plot diagram for the size of bag-of-terms in experimental dataset}. Source: Agarwal et al. \cite{agarwal2015using}}
 \label{bow}
\end{minipage}
\end{figure*}
As an academic researcher, we believe and encourage academic code or software sharing in the interest of improving openness and research reproducibility. We release our term obfuscation detection tool \textit{Parikshan} in public domain so that other researchers can validate our scientific claims and use our tool for comparison or benchmarking purposes. \textit{Parikshan} is a proof-of-concept hosted on GitHub which is a popular web-based hosting service for software development projects. We provide installation instructions and a facility for users to download the software as a single zip-file. Another reason of hosting on GitHub is due to an integrated issue-tracker which makes reporting issues easier by our users (and also GitHub facilitates easier collaboration and extension through pull-requests and forking). The link to \textit{Parikshan} on GitHub is: https://github.com/ashishsureka/Parikshan. We believe our tool has utility and value in the domain of intelligence and security informatics and in the spirit of scientific advancement, select GPL license (restrictive license) so that our tool can never be closed-sourced.
\subsection{Experimental Dataset}\label{section:exp_dataset}
We conduct experiments on publicly available dataset so that our results can be used for comparison and benchmarking. We download two datasets: Enron e-mail corpus\footnote{\url{http://verbs.colorado.edu/enronsent/}} and Brown news corpus\footnote{\url{http://www.nltk.org/data.html}}. We also use the examples extracted from $4$ research papers on word substitution. Hence we have a total of three experimental datasets to evaluate our proposed approach. We believe conducting experiments on three diverse evaluation dataset will prove the generalizability of our approach and thus strengthen the conclusions. Enron e-mail corpus consists of about half a million e-mail messages sent or received by about 158 employees of Enron corporation. This dataset was collected and prepared by the CALO Project\footnote{\url{https://www.cs.cmu.edu/\~./enron/}}. We perform a random sampling on the dataset and select $9000$ unique sentences for substitution. Brown news corpus consists of about a million words from various categories of formal text and news (for example, political, sports, society and cultural). This dataset was created in 1961 at Brown University. Since the writing style in Brown news corpus is much more formal than Enron e-mail corpus, we use these two different datasets to examine the effectiveness of our approach.\\
\indent We perform a word substitution technique (refer to Section \ref{section:substitution_technique}) on a sample of $9000$ sentences from Enron e-mail corpus and all $4600$ of Brown news corpus. Figure \ref{dataset_stats} shows the statistics of both the datasets before and after the word substitution. Figure \ref{stats_brown} and \ref{stats_enron} also illustrates the variation in number of sentences substituted using traditional approach (proposed in Fong et. al. \cite{Fong08}) and our approach. Figure \ref{stats_brown} and \ref{stats_enron} reveals that COCA is a huge corpus and has more nouns in the frequency list in comparison to BNC frequency list. Table \ref{tab:stats_brown_enron} displays the exact values for the points plotted in the two bar charts of Figure \ref{dataset_stats}. \\
\begin{table*}[t]
\normalsize
\sffamily
\centering
\caption{\textbf{Experimental Dataset Statistics for the Brown News Corpus (BNC) and Enron Mail Corpus (EMC) (Refer to Figure \ref {dataset_stats} for the Graphical Plot of the Statistics), \#=Number of}. Source: Agarwal et al. \cite{agarwal2015using}} \label{tab:stats_brown_enron}
\begin{tabular}{lp{11cm}ll}
Abbr & Description & BNC & EMC \\ 
\hline
Corpus & Total sentences in brown news corpus & $4607$ & $9112$ \\ 
5-15 & Sentences that has length between 5 to 15 & $1449$ & $2825$ \\ 
N-BNC & Sentences that has their first noun in BNC (british national corpus) & $2214$ & $3587$ \\ 
N-COCA & Sentences that has their first noun in 100 K list (COCA) & $2393$ & $4006$\\ 
N-H-W & If first noun has an hypernym in WordNet & $3441$ & $5620$\\ 
En-BNC & English sentences according to BNC & $2146$ & $3430$ \\ 
En- Java & English sentences according to Java language detection library & $4453$ & $8527$ \\ 
S'-BNC & \#Substituted sentences using BNC list & $2146$ & $3430$ \\ 
S'-COCA & \#Substituted sentences using COCA (100K) list & $2335$ & $3823$ \\ 
S'-B-5-15 & \#Substituted sentences (between length of 5 to 15) using BNC list & $666$ & $1051$\\ 
S'-C-5-15 & \#Substituted sentences (between length of 5 to 15) using COCA list & $740$ & $1191$ \\ 
\end{tabular}
\end{table*}
\begin {table*}[t]
\sffamily
\normalsize
\centering
\caption{\textbf{Concrete Examples of Sentences Presented in EMC and BNC Corpus Discarded While Word Substitution}. Source: Agarwal et al. \cite{agarwal2015using}} \label{discarded} 
\begin{center}
 \begin{tabular}{lp{.6\textwidth}p{.25\textwidth}}
Corpus&Sentence&Reason\\ 
\hline
EMC & Since we're ending 2000 and going into a new sales year I want to make sure I'm not holding resource open on any accounts which may not or should not be on the list of focus accounts which you and your team have requested our involvement with. & Sentence length is not between 5 to 15\\ 
EMC & next \underline{Thursday} at 7:00 pm Yes yes yes. & First noun is not in BNC/COCA list\\ 
BNC & The City Purchasing Department the jury said is lacking in experienced clerical personnel as a result of city personnel policies & Sentence length is not between 5 to 15\\ 
BNC & \underline{Dr} Clark holds an earned Doctor of Education degree from the University of Oklahoma & First noun does not have a hypernym in WordNet\\ 
\end{tabular}
\end{center}
\end{table*}
Table \ref{tab:stats_brown_enron} reveals that for Brown news corpus, using BNC (British National Corpus) frequency list we are able to detect only $2146$ English sentences while using Java language detection library we are able to detect $4453$ English sentences. Similarly, in Enron e-mail corpus, BNC frequency list detects only $3430$ English sentences while Java language detection library identifies $8527$ English sentences. Therefore using COCA frequency list and Java language detection library we are able to substitute more sentences ($740$ and $1191$) in comparison to previous approach ($666$ and $1051$). Table \ref{tab:stats_brown_enron} reveals that initially we have a dataset of $4607$ and $9112$ sentences for BNC and EMC respectively. After word substitution we are remaining with only $740$ and $1191$ sentences. Some sentences are discarded because they do not satisfy several conditions of word obfuscation. Table \ref{discarded} shows some concrete examples of such sentences from BNC and EMC datasets.\\
\indent We use $740$ substituted sentences from Brown news corpus, $1191$ sentences from Enron e-mail corpus and $22$ examples from previous research papers as our testing dataset. As shown in research framework (refer to Figure \ref{fig:Step-A-B}) we apply a part-of-speech tagger on each sentence to remove non-content bearing terms. Figure \ref{pos_tags} illustrates the frequency of common part-of-speech tags present in Brown news corpus (BNC) and Enron e-mail corpus (EMC). As shown in Figure \ref{pos_tags}, the most frequent part-of-speech in the dataset is nouns followed by verbs. Figure \ref{bow} shows the length of bag-of-terms for every sentence present in BNC and EMC datasets. Figure \ref{bow} reveals that $5$ sentences in Enron e-mail corpus and $6$ sentences in Brown news corpus have an empty bag-of-terms which makes the system difficult to identify an obfuscated term. Figure \ref{bow} reveals that for majority of sentences size of bag-of-terms varies between $2$ to $6$. It also illustrates the presence of sentences that have insufficient number of concepts (size \textless $2$) or the sentences that have large number of concepts (size \textgreater $7$).
\begin{algorithm}[t]
\normalsize
\SetAlgoLined\DontPrintSemicolon
\KwData{Sentence $S$, Frequency List $COCA$, WordNet DataBase $W_{DB}$}
\KwResult{Substituted Sentence $S'$}
\nl \If{($5<S.length < 15$)}{
\nl $tokens \leftarrow $S.tokenize()\;
\nl $POS \leftarrow $S.pos\_tag()\;
\nl $NF \leftarrow $token[POS.indexOf($"NN"$)]\;
\nl \If{($COCA.has(NF)$ AND $W_{DB}.has(NF.hypernym)$)}{
\nl $lang \leftarrow $ S.Language Detection\;
\nl \If {($lang == "en"$)}{
\nl $F_{NF} \leftarrow $COCA.freq(NF)\;
\nl $F_{NF'} \leftarrow $ COCA.nextHigherFreq($F_{NF}$)\;
\nl $ NF' \leftarrow $ COCA.hasFrequency($F_{NF'}$)\;
\nl $S' \leftarrow $ S.replaceFirst(NF, NF')\;
\nl return S'\;
}}}
\caption{Text Substitution Technique}\label{algo:substitution}
\end{algorithm}
\begin {table*}[ht!]
\sffamily
\normalsize
\caption {\textbf{Example of Term Substitution using COCA Frequency List. NF= First Noun/ Original Term, ST= Substituted Term}. Source: Agarwal et al. \cite{agarwal2015using}} \label{data_in_subtitution} 
\begin{center}
 \begin{tabular}{p{.26\textwidth}llllp{.26\textwidth}}
 Sentence&NF&Freq&ST&Freq&Sentence\\ 
 \hline
Any opinions expressed herein are solely those of the author. & Author & 53195 & Television & 53263 & Any opinions expressed herein are solely those of the television.\\
What do you think that should help you score women. & Score & 17415 & Struggle & 17429 & What do you think that should help you struggle women.\\
This was the coolest calmest election I ever saw Colquitt Policeman Tom Williams said & Election & 40513 & Republicans & 40515& This was the coolest calmest republicans I ever saw Colquitt Policeman Tom Williams said\\
The inadequacy of our library system will become critical unless we act vigorously to correct this condition & Inadequacy & 831 & Inevitability & 831 & The inevitability of our library system will become critical unless we act vigorously to correct this condition\\
\end{tabular}
\end{center}
\end{table*}
\begin {table*}[ht]
\sffamily
\small
\caption{\textbf{List of Original and Substituted Sentences used as Examples in Papers on Word Obfuscation in Adversarial Communication}. Source: Agarwal et al. \cite{agarwal2015using}} \label{tab:origsubsprevwork} 
\begin{center}
 \begin{tabular}{cp{7cm}p{6.8cm}l}
& Original Sentence & Substituted Sentence & Result \\ 
\hline
1 & the \underline{bomb} is in position \cite{Fong06} & the \underline{alcohol} is in position & alcohol \\ 
2 & \underline{copyright} 2001 south-west airlines co all rights reserved \cite{Fong06} & \underline{toast} 2001 southwest airlines co all rights reserved & southwest \\ 
3 & please try to maintain the same \underline{seat} each class \cite{Fong06} & please try to maintain the same \underline{play} each class & try \\ 
4 & we expect that the \underline{attack} will happen tonight \cite{Fong08} & we expect that the \underline{campaign} will happen tonight & campaign \\ 
5 & an \underline{agent} will assist you with checked baggage \cite{Fong08} & an \underline{vote} will assist you with checked baggage & vote \\ 
6 & my \underline{lunch} contained white tuna she ordered a parfait \cite{Fong08} & my \underline{package} contained white tuna she ordered a parfait & package \\ 
7 & please let me know if you have this \underline{information} \cite{Fong08} & please let me know if you have this \underline{men} & know \\ 
8 & It was one of a \underline{series} of recommendations by the Texas Research League \cite{Fong08} & It was one of a \underline{bank} of recommendations by the Texas Research League & recomm. \\ 
9 & The \underline{remainder} of the college requirement would be in general subjects \cite{Fong08} & The \underline{attendance} of the college requirement would be in general subjects & attendance \\ 
10 & A \underline{copy} was released to the press \cite{Fong08} & An \underline{object} was released to the press & released \\ 
11 & works need to be done in \underline{Hydrabad} \cite{Deshmukh14} & works need to be done in \underline{H} & H \\ 
12 & you should arrange for a preparation of \underline{blast} \cite{Deshmukh14} & you should arrange for a preparation of \underline{daawati} & daawati \\ 
13 & my friend will come to deliver you a \underline{pistol} \cite{Deshmukh14} & my friend will come to deliver you a \underline{CD} & CD \\ 
14 & collect some people for work from \underline{Gujarat} \cite{Deshmukh14} & collect some people for work from \underline{Musa} & Musa \\ 
15 & you will find some \underline{bullets} in the bag \cite{Deshmukh14} & you will find some \underline{pen drives} in the bag & pen drives \\ 
16 & come at \underline{Delhi} for meeting \cite{Deshmukh14} & come at \underline{Sham} for meeting & Sham \\ 
17 & send one person to \underline{Bangalore} \cite{Deshmukh14} & send one person to \underline{Bagu} & Bagu \\ 
18 & Arrange some \underline{riffles} for next operation \cite{Deshmukh14} & Arrange some \underline{DVDs} for next operation & DVDs\\ 
19 & preparation of \underline{blast} will start in next month \cite{Deshmukh14} & preparation of \underline{Daawati} work will start in next month & Daawati \\ 
20 & find one place at \underline{Hydrabad} for operation \cite{Deshmukh14} & find one place at \underline{H} for operation & H \\ 
21 & He remembered sitting on the wall with a cousin, watching the German \underline{bomber} fly over \cite{Jabbari08} & He remembered sitting on the wall with a cousin, watching the German \underline{dancers} fly over & German \\ 
22 & Perhaps no ballet has ever made the same impact on \underline{dancers} and audience as Stravinsky's "Rite of Spring \cite{Jabbari08} & Perhaps no ballet has ever made the same impact on \underline{bomber} and audience as Stravinsky's "Rite of Spring & bomber \\ 
\end{tabular}
\end{center}
\end{table*}
\subsection{Term Substitution Technique}\label{section:substitution_technique}
We substitute a term in a sentence using an adaptive version of a substitution technique originally proposed by Fong et. al. \cite{Fong08}. Algorithm \ref{algo:substitution} describes the steps to obfuscate a term in a given sentence. We use WordNet database\footnote{\url{http://wordnet.princeton.edu/wordnet/download/}} as a language resource and the Corpus of Contemporary American English (COCA)\footnote{COCA is a corpus of Americal English that contains more than $450$ million words collected from 1990-2012. http://www.wordfrequency.info/} as a word frequency data. In Step $1$, we check the length of a given sentence $S$. If the length is between $5$ to $15$ then we proceed further otherwise we discard that sentence. In Steps $2$ and $3$, we tokenize the sentence $S$ and apply part-of-speech tagger to annotate each word. In Step $4$, we identify the first noun $NF$ from this word sequence $POS$. In Steps $5$, we check if $NF$ is present in COCA frequent list and has an hypernym in WordNet. If the condition satisfies then we detect the language of the sentence using Java language detection library\footnote{\url{https://code.google.com/p/language-detection/}}. If the sentence language is not English then we ignore it and if it is English then we further process it. In Steps $8$ to $11$, we check the frequency of $NF$ in COCA corpus and replace the term in the sentence by a new term $NF'$ with the next higher frequency in COCA frequency list. This new term $NF'$ is the obfuscated term. If $NF$ has the highest frequency in COCA corpus then we substitute it with the term which appears immediate before $NF$ in frequency list. If two terms have the same frequency then we sort those terms in alphabetical order and select immediate next term to $NF$ for substitution. Table \ref{data_in_subtitution} shows some concrete examples of substituted sentences taken from Brown news corpus and Enron e-mail corpus. In Table \ref{data_in_subtitution}, Freq denotes the frequency of first noun and it's substituted term in COCA frequency list. Table \ref{data_in_subtitution} also shows an example where two terms have the same frequency. We replace the first noun with the term that has equal frequency and is next immediate to $NF$ in alphabetical order.\\
\indent In Fong et. al.; they use British National Corpus (BNC) as word frequency list. We replace BNC list by COCA frequency list because it is the largest and most accurate frequency data of English language and is 5 times bigger than the BNC list. The words in COCA are divided among a variety of texts (for example, spoken, newspapers, fiction and academic texts) which are best suitable for working with common sense knowledge base. In Fong et. al; they identify the sentence to be in English language if $NF$ is present in BNC frequency list. Since the size of BNC list is comparatively small, we use Java language detection library for identifying the language of the sentence \cite{nakatani2010langdetect}. Java language detection library supports $53$ languages and is much more flexible in comparison to BNC frequency list. 
\subsection{Experimental Results}
\subsubsection{Examples from Research Papers (ERP)}
\begin{table}[ht!]
\sffamily
\normalsize
\caption {\textbf{Accuracy Results for Brown News Corpus (BNC) and Enron Mail Corpus (EMC)}. Source: Agarwal et al. \cite{agarwal2015using}} \label{tab:accuracy_results} 
\begin{center}
 \begin{tabular}{lp{0.1\textwidth}p{0.1\textwidth}p{0.1\textwidth}l}
&Total Sentences&Correctly Identified&Accuracy Results&NA \\ 
 \hline
BNC & 740 & 573 & 77.4\% & 46\\ 
EMC & 1191 & 629 & 62.9\% & 125\\ 
\end{tabular}
\end{center}
\end{table}
As described in section \ref{section:exp_dataset}, we run our experiments on examples used in previous papers. Table \ref{tab:origsubsprevwork} shows $22$ examples extracted from $4$ research papers on term obfuscation (called as ERP dataset). Table \ref{tab:origsubsprevwork} shows the original sentence, substituted sentence, research paper and the result produced by our tool. Experimental results reveal $72.72\%$ accuracy of our solution approach ($16$ out of $22$ correct output).
\begin {table}[t]
\sffamily
\normalsize
\caption {\textbf{Concrete Examples of Sentences with Size of Bag-of-terms (BoT) Less Than 2}. Source: Agarwal et al. \cite{agarwal2015using}} \label{tab:bow_less_2} 
\begin{center}
 \begin{tabular}{p{0.32\textwidth}l}
Corpus: Sentence& BoT: Size\\ 
\hline
BNC: That was before I studied both & []: 0\\ 
BNC: The jews had been expected & [jews]: 1\\ 
BNC: if we are not discriminating in our cars & [car]: 1\\ 
EMC: What is the benefits? & [benefits]: 1\\ 
EMC: Who coined the adolescents? & [adolescents]: 1\\ 
EMC: Can you help? his days is 011 44 207 397 0840 john& [day]: 1\\ 
\end{tabular}
\end{center}
\end{table}
\begin {table*}[t]
\sffamily
\normalsize
\caption{\textbf{Concrete Examples of Sentences with the Presence of Technical Terms and Abbreviations.} Source: Agarwal et al. \cite{agarwal2015using}} \label{tab:tech_abbr}
\begin{center}
 \begin{tabular}{p{.65\textwidth}ll}
Sentence&Tech Terms&Abbr\\ 
\hline
\#4. artifacts 2004-2008 maybe 1 trade a day. & Artifacts & -\\ 
We have put the interview on IPTV for your viewing pleasure.& Interview, IPTV & IPTV\\ 
Will talk with KGW off name. & - & KGW\\ 
We are having males backtesting Larry May's VaR. & backtesting & VAR\\ 
Internetworking and today American Express has surfaced. & Internetworking & - \\ 
I do not know their particles yet due to the Enron PRC meeting conflicts. & Enron & PRC \\ 
The others may have contracts with LNG consistency owners. & - & LNG \\ 
\end{tabular}
\end{center}
\end{table*}
\begin {table*}[t]
\sffamily
\normalsize
\caption{\textbf{Concrete Examples of Long Sentences (Length of Bag-of-terms \textgreater = 5) Where Substituted Term is Identified Correctly.} Source: Agarwal et al. \cite{agarwal2015using}} \label{tab:bow5} 
\begin{center}
\begin{tabular}{lp{.43\textwidth}lp{.30\textwidth}}
Corpus&Sentence&Original&Bag-of-Terms\\
\hline
\textbf{BNC}&He further proposed grants of an unspecified \underline{input} for experimental hospitals & Sum & [grants, unspecified, input, experimental, hospitals]\\
\textbf{BNC}&When the gubernatorial \underline{action} starts Caldwell is expected to become a campaign coordinator for Byrd& Campaign & [gubernatorial, action, Caldwell, campaign, coordinator, Byrd]\\
\textbf{BNC}&The entire \underline{arguments} collection is available to patrons of all members on interlibrary loans &Headquarters &[entire, argument, collection, available, patron, member, interlibrary, loan]\\
\textbf{EMC}&Methodologies for accurate skill-matching and \underline{pilgrims} efficiencies=20 Key Benefits ?& Fulfillment & [methodologies, accurate, skill, pilgrims, efficiencies, benefits]\\
\textbf{EMC}&PERFORMANCE REVIEW The \underline{measurement} to provide feedback is Friday November 17.& Deadline & [performance, review, measurement, feedback, friday, november]\\
\end{tabular}
\end{center}
\end{table*}
\begin{figure*}[ht!]
\centering
\begin{minipage}{.40\textwidth}
 \includegraphics[width=0.9\textwidth]{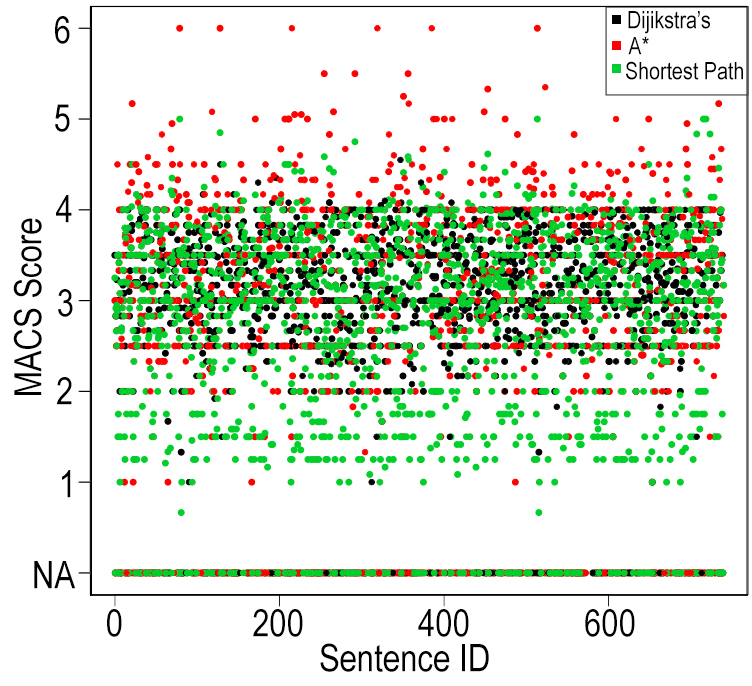}
 \caption{MAC Score of concepts for each sentence for Brown news corpus}
 \label{fig:brown}
 \end{minipage}%
\qquad
\begin{minipage}{.40\textwidth}
 \includegraphics[width=0.9\textwidth]{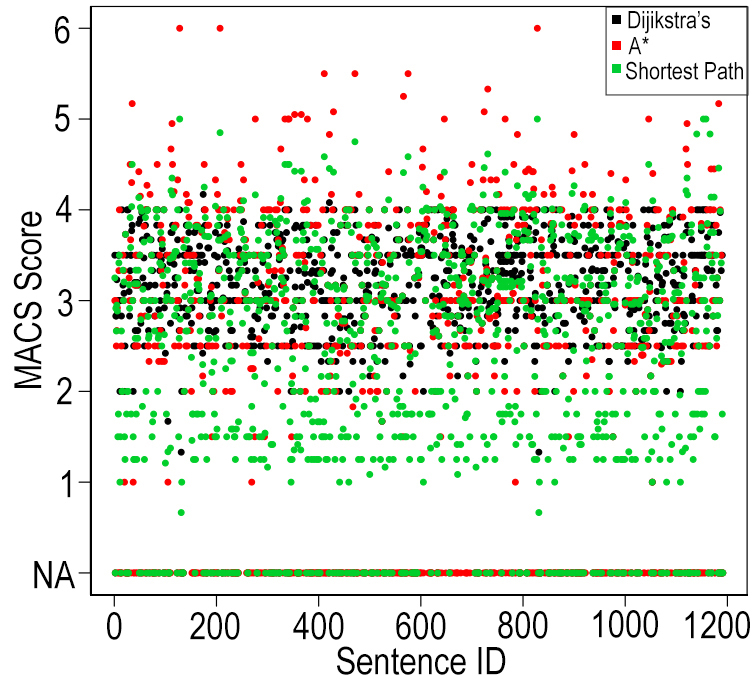}
 \caption{MAC Score of concepts for each sentence for Enron mail corpus}
 \label{fig:enron}
 \end{minipage}
\end{figure*}
\begin{figure*}[ht!]
\centering
\begin{minipage}{.40\textwidth}
 \includegraphics[width=0.8\textwidth]{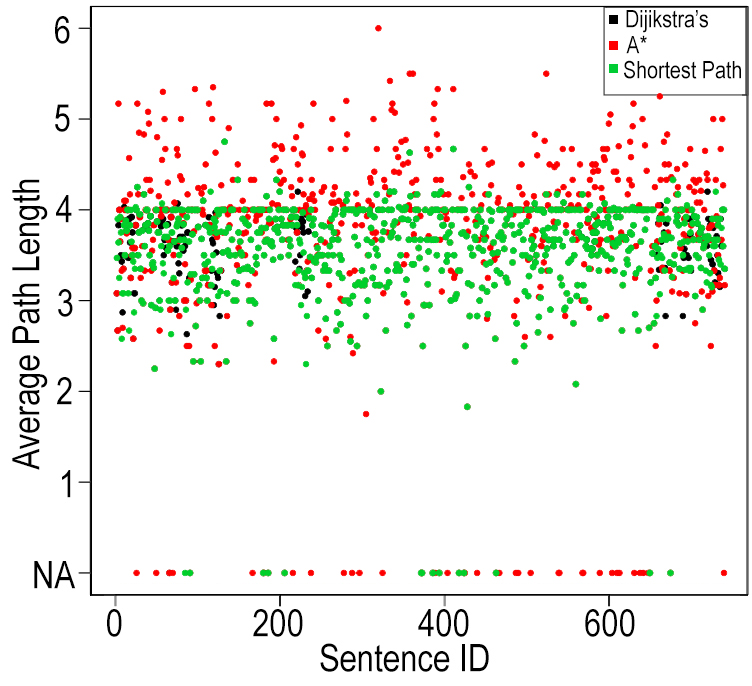}
 \caption{Average path length of concepts for each sentence for Brown news corpus}
 \label{avg_brown}
 \end{minipage}%
\qquad
\begin{minipage}{.40\textwidth}
 \centering
 \includegraphics[width=0.8\textwidth]{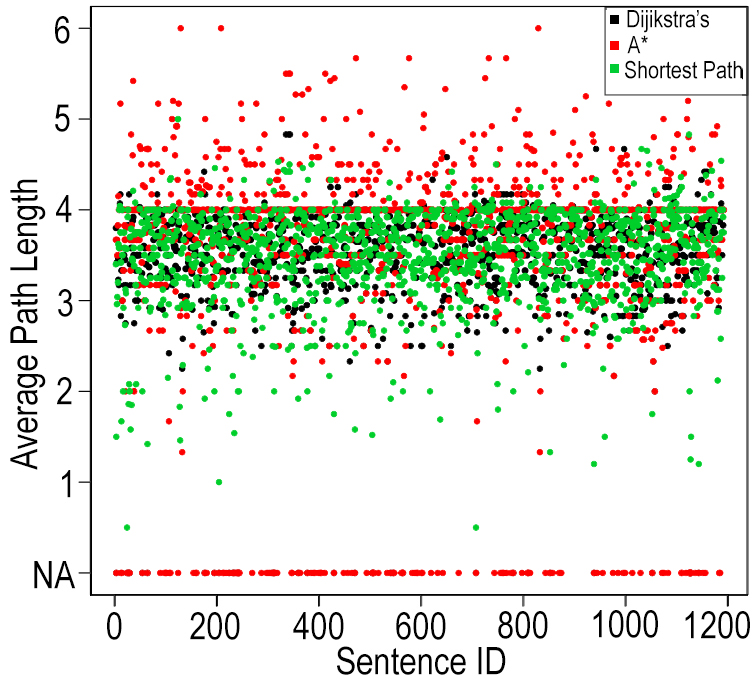}
 \caption{Average path length of concepts for each sentence for Enron mail corpus}
 \label{avg_enron}
 \end{minipage}
\end{figure*}
\subsubsection{Brown News Corpus (BNC) and Enron Email Corpus (EMC)}
To evaluate the performance of our solution approach we collect results for all $740$ and $1191$ sentences from BNC and EMC datasets respectively. Table \ref{tab:accuracy_results} reveals an accuracy of $77.4$\% ($573$ out of $740$ sentences) for BNC and an accuracy of $62.9$\% ($629$ out of $1191$ sentences) for EMC. "NA" denotes the number of sentences where the concepts present in bag-of-terms are not good enough to identify an obfuscated term (bag-of-terms length \textless $2$). Table \ref{tab:bow_less_2} shows some concrete examples of these sentences from BNC and EMC datasets. Table \ref{tab:accuracy_results} also reveals that for BNC dataset our tool outperforms the EMC dataset with a difference of $14.5$\% in overall accuracy. The reason behind this major fall in the accuracy is that Enron e-mails are written in much more informal manner and length of bag-of-terms for those sentences is either too small (\textless 2) or too large (\textgreater 6). Also the sentences generated from these e-mails contain several technical terms and abbreviations. These abbreviations are annotated as nouns in part-of-speech tagging and do not exist in common sense knowledge-base. Table \ref{tab:tech_abbr} shows some concrete examples of such sentences. Table \ref{tab:tech_abbr} also reveals that there are some sentences that contain both abbreviations and technical terms. Experimental results reveals that our approach is effective and able to detect obfuscated term correctly in long sentences containing more than 5 concepts in bag-of-terms. Table \ref{tab:bow5} shows some examples of such sentences present in BNC and EMC datasets.\\
\indent We believe that our approach is more generalized in comparison to existing approaches. Word obfuscation detection techniques proposed by Deshmukh et al. \cite{Deshmukh14} Fong et. al. \cite{Fong08} and Jabbari et al.\cite{Jabbari08} are focused towards the substitution of first noun in a sentence. The bag-of-term approach is not limited to the first noun of a sentence. We use a bag-of-terms approach that is able to identify any term that has been obfuscated.\\
\indent \textbf{Minimum Average Conceptual Similarity (MACS) Score:} Figures \ref{fig:brown} and \ref{fig:enron} shows the minimum average conceptual similarity (MACS) score for Brown news corpus and Enron e-mail corpus respectively. Figure \ref{fig:brown} also reveals that using Dijikstra's algorithm, majority of the sentences have mean average path length between $2$ and $3.5$. For Shortest path algorithm one third of sentences have mean average path length between $1$ and $2$. That means in shortest path metrics, we find many directly connected edges. In Figure \ref{fig:brown}, we also observe that for half of sentences Dijikstra's and shortest path algorithms have similar MACS score. If two concepts are not reachable then we use $4$ as a default value for no-path. MACS score between $4.5$ to $6$ shows the absence of path among concepts or a relatively much longer path in the knowledge-base. Figure \ref{fig:brown} reveals that for some sentences A* algorithm has a mean average path length between $4$ and $6$. Figure \ref{fig:enron} illustrates that using A* and Dijikstra's algorithm, majority of sentences have a mean average path length between $3$ to $4$. It shows that for many sentences bag-of-terms have concepts that are conceptually un-related. This happens because Enron e-mail corpus has many technical terms that are not semantically related to each other in commonsense knowledge-base. Similar to Brown news corpus, we observe that for half of sentences shortest path algorithm has mean average path length between $1$ and $2$. In comparison to Brown news corpus, for Enron e-mail corpus A* algorithm has large MACS score for very few sentences. It reveals that either the concepts are connected by one or two concepts in between or they are not connected at all (no-path).\\
\indent \textbf{Average Path Length Score:} Figures \ref{avg_brown} and \ref{avg_enron} shows the average path length between concepts for each sentence present in the BNC and EMC datasets respectively. Figure \ref{avg_enron} reveals that for Dijikstra's and shortest path algorithms, $80$\% sentences of brown news corpus have same average path length. Also majority of sentences have an average path length between $2.5$ and $3.5$. Similar to Figures \ref{fig:brown} and \ref{fig:enron} "NA" denotes the sentences with insufficient number of concepts. Figure \ref{avg_brown} also reveals the presence of obfuscated term in the sentence. Since no sentence has an average length of $1$ and similarly, only $1$ sentence has an average length of $2$. This implies the presence of terms that are not conceptually related to each other. Figure \ref{avg_enron} shows that majority of sentences have average path length between $2.5$ and $4$ for all three distance metrics. Figure \ref{avg_enron} also reveals that for some sentences shortest path algorithm has average path length between $0.5$ and $2$. Figure \ref{avg_enron} shows that for some sentences all three algorithms have average path length between $4$ and $6$. This happens because of the presence of a few technical terms and abbreviations. These terms have no path in ConceptNet 4.0 and therefore assigned a default value of 4.0 which increases the average path length for whole bag-of-terms.
\section{Threats to Validity and Limitations}
The proposed solution approach for textual or term obfuscation detection uses ConceptNet knowledge-base for computing conceptual and semantic similarity between any two given terms. We use version $4.0$ of ConceptNet and the solution result is dependent on the nodes and the relationships between the nodes in the specific version of the ConceptNet knowledge-base. Hence a change in the version of the ConceptNet may have some effect on the outcome. For example, the number of paths between any two given concepts or the number of edges in the shortest path between any two given concepts may vary from one version to another. One of the limitations of our approach is that as the size of the bag-of-terms increases, the number of times the function to compute shortest path between two nodes (and hence the overall computation time) increases substantially. 
\section{Conclusions}
We present an approach to detect term obfuscation in adversarial communication using ConceptNet common-sense knowledge-base. The proposed solution approach consists of identifying the out-of-context term in a given sentence by computing conceptual similarity between the terms in the given sentence. We compute the accuracy of the proposed solution approach on three test datasets: example sentences from research papers, brown news corpus and email news corpus. Experimental results reveal an accuracy of $72.72\%$, $77.4\%$ and $62.0\%$ respectively on the three dataset. Empirical evaluation and validation shows than the proposed approach is effective (an average accuracy of more than $70\%$) for the task of identifying obfuscated term in a given sentence. Experimental results demonstrate that our approach is also able to detect term obfuscation in long sentences containing more than $5-6$ concepts. Furthermore, we demonstrate that the proposed approach is generalizable as we conduct experiments on nearly $2000$ sentences belonging two three different datasets and diverse domains. 
\bibliographystyle{IEEEtran}
\bibliography{ICDCIT}
\end{document}